\documentclass[preprintnumbers,amsmath,amssymb,floatfix,11pt,prd,superscriptaddress,nofootinbib]{revtex4}
\usepackage{graphicx}
\usepackage{epsfig}
\usepackage{amssymb, amscd, amsmath,amsthm,amsfonts,enumerate}
\usepackage{bm}
\usepackage{amsfonts}
\usepackage{color}
\begin{document}

\title{Cosmic accelerated expansion and the entropy-corrected holographic dark energy}

\author{\textbf{H. Mohseni Sadjadi}}
\email{mohsenisad@ut.ac.ir} \affiliation{Department of Physics,
University of Tehran, P.O.B. 14395-547, Tehran, Iran}

\author{\textbf{Mubasher Jamil}}
\email{mjamil@camp.nust.edu.pk,jamil.camp@gmail.com}
\affiliation{Center for Advanced Mathematics and Physics, National
University of Sciences and Technology, H-12, Islamabad, Pakistan}

\begin{abstract}
\vspace*{1.5cm} \centerline{\bf Abstract} \vspace*{.5cm} By
considering the logarithmic correction to the energy density, we
study the behavior of Hubble parameter in the holographic dark
energy model.  We assume that the universe is dominated by
interacting dark energy and matter and the accelerated expansion of
the universe, which may be occurred in the early universe or late
time, is studied.
\end{abstract}

\maketitle

\newpage

\section{Introduction}
To explain the cosmic accelerated expansion of the universe
\cite{Rie}, and motivated by the holographic principle
\cite{Suss1}, a model of dark energy has been proposed
\cite{Coh,Hsu,Li,Huang,HDE,Setare1,wang0} which has been tested
and constrained by various astronomical observations \cite{Xin}.
This proposal is generically known as the ``Holographic Dark
Energy" (HDE). Its definition is originally extracted from the
entropy-area relation which depends on the theory of gravity. In
the thermodynamics of black hole horizons, there is a maximum
entropy in a box of length $L$, commonly termed, the
Bekenstein-Hawking entropy bound, $S\simeq m_P^2L^2$, which scales
as the area of the box $A \sim L^2$ rather than the volume $V \sim
L^3$. Here $m^2_p =(8\pi G)^{-1}$ is the reduced Planck mass. In
this context, Cohen et al. \cite{Coh} proposed that in quantum
field theory a short distance cutoff $\Lambda$ is related to a
long distance cutoff $L$ due to the limit set by formation of a
black hole, which results in an upper bound on the zero-point
energy density. In line with this suggestion, Hsu and Li
\cite{Hsu,Li} argued that this energy density could be viewed as
the holographic dark energy density satisfying $\rho_{d }
=3n^2m^2_p/L^2$, where $L$ is the size of a region which provides
an IR cut-off, and the numerical constant $3n^2$ is introduced for
convenience.

It is essential to notice that in the literature, various scenarios
of HDE have been studied via considering different system's IR
cutoff. In the absence of interaction between dark matter and dark
energy in flat universe, Li \cite{Li} discussed three choices for
the length scale $L$ which is supposed to provide an IR cutoff. The
first choice is the Hubble radius, $L=H^{-1}$ \cite{Hsu}, which
leads to a wrong equation of state for dark energy ($\omega=0$),
namely that for dust. The second option is the particle horizon
radius. In this case it is impossible to obtain an accelerated
expansion. Only the third choice, the identification of $L$ with the
radius of the future event horizon gives the desired result, namely
a sufficiently negative equation of state to obtain an accelerated
universe. However, as soon as an interaction between dark energy and
dark matter is taken into account, the first choice, $L=H^{-1}$, in
flat universe, can simultaneously drive accelerated expansion and
solve the coincidence problem \cite{pav1}. It was also demonstrated
that in the presence of an interaction, in a non-flat universe, the
natural choice for IR cutoff could be the apparent horizon radius
\cite{shey1}.

As we earlier mentioned that the black hole entropy $S$ plays a
central role in the derivation of HDE. Indeed, the definition and
derivation of holographic energy density  ($\rho_{d }
=3n^2m^2_p/L^2$) depends on the entropy-area relationship $S\sim
 A \sim L^2$ of black holes in Einstein's gravity, where $A \sim L^2$
represents the area of the horizon. However, this definition can
be modified from the inclusion of the effects of  thermal
fluctuations around equilibrium \cite{therm}, quantum fluctuations
\cite{quant}, or by considering the loop quantum gravity (LQG)
\cite{Rovelli}, all leading almost to the same result. The
corrections provided to the entropy-area relationship leads to the
curvature correction in the Einstein-Hilbert action and vice versa
\cite{Zhu}. The corrected entropy takes the form \cite{modak}
\begin{equation}\label{S}
S=\frac{A}{4G}+\gamma \ln {\frac{A}{4G}}+\delta,
\end{equation}
where $\gamma$ and $\delta$ are dimensionless constants of order
unity. The exact values of these constants are not yet determined
and still debatable. Taking the corrected entropy-area relation
(\ref{S}) into account, the energy density of the HDE will be
modified as well. On this basis, Wei \cite{wei} proposed the
energy density of the so-called ``entropy-corrected holographic
dark energy''
 (ECHDE) in the form
\begin{equation}\label{rhoS}
\rho _{d }=3n^2m_{p}^{2}L^{-2}+\gamma L^{-4}\ln
(m_{p}^{2}L^{2})+\delta L^{-4}.
\end{equation}
In the special case  $\gamma=\delta=0$, the above equation yields
the well-known holographic energy density.

To solve some essential problems in standard cosmology, it is
believed that there was also another stage of accelerated
expansion in the early universe known as inflation \cite{inf}. In
the same way that some models, proposed to explain the inflation(
such as a scalar field in slow roll model \cite{slow}), have been
used to explain the present acceleration of the universe,  one can
examine different models of dark energy, to explain the evolution
of the universe in the inflation era.

In this paper we study all possible behaviors of the Hubble
parameter in a universe dominated by (ECHDE) and a barotopic
matter and study some consequences of this model in accelerated
expansion in late time and also in the inflation era.

The plan of the paper is as follows: In section 2 we construct a
cosmological model and derive some useful expressions for our
further uses. In section 3, we obtain different possible behaviors
of Hubble parameter in the interacting ECHDE model.  In section 4
we discuss the possible implications of this model in the
accelerated expansion of the universe in the early stage and in
the present epoch separately(we do not intend to study the whole
history of the universe in a same framework). We also check the
conditions under which the universe will undergo multiple
acceleration-deceleration phases. Finally we conclude this paper
in the last section. We use units $\hbar=G=c=k_B=1$.

\section{The model}
We consider a spatially flat Friedman Robertson walker (FRW)
universe dominated by dark energy and matter (this can be dark
matter, radiation and so on). The Friedman equations read
\begin{eqnarray}\label{1}
H^2&=&{8\pi\over 3}(\rho_d+\rho_m)\nonumber \\
\dot{H}&=&-4\pi(P_d+\rho_d+\rho_m),
\end{eqnarray}
where $\rho_d$ and $\rho_m$ are densities of dark energy and
matter respectively. $P_d$ is the pressure of dark energy and the
Hubble parameter, $H$, is assumed to be differentiable. As we have
noticed in the introduction, the corrections (\ref{rhoS}), can be
obtained from the loop quantum gravity, as well as by the
inclusion of the effects of thermal fluctuations around
equilibrium , quantum fluctuations, or by considering charge or
mass fluctuations. So it is safe to study this problem in the
framework of FRW cosmology, using ordinary Friedman equations as
was mentioned in \cite{wei}.

The matter and dark energy are allowed to exchange
energy\cite{interaction} via the source term $Q$:
\begin{eqnarray}\label{2}
\dot{\rho}_d+3H(w_d+1)\rho_d&=&-Q\nonumber \\
\dot{\rho}_m+3H(1+w_m)\rho_m&=&Q.
\end{eqnarray}
Because of this interaction term, we have not the conservation of
partial stress-energy tensors of matter and dark energy :$T^{\mu
\nu}_{(matter);\mu}=-T^{\mu \nu}_{(dark);\mu}\neq 0$. It is
assumed that the matter component and dark energy have the same
velocity which is the velocity of the whole fluid, $V$. We can
write
\begin{equation}\label{r1}
T^{\mu \nu}_{(matter);\mu}V_\nu=-T^{\mu \nu}_{(dark);\mu}V_\nu=0.
\end{equation}
In the scalar field model of inflation, after the slow roll
regime, the scalar field, whose energy density is $\rho_d$, decays
to radiation via a rapid coherent oscillation. The source term for
this decay which allows the radiation creation from inflaton is
taken as \cite{Kolb}
\begin{equation}\label{r2}
Q=\alpha H \rho_d,
\end{equation}
where $\alpha$ is a constant. Also, in dark energy models
different interactions between matter and dark energy are assumed.
As the nature of dark energy has not yet been known, these
interactions are taken from other models such as string theory and
scalar tensor theory and so on, or as
\begin{eqnarray}\label{r3}
Q&=&\beta H (\rho_d+\rho_m)\nonumber \\
Q&=&\eta H\rho_m,
\end{eqnarray}
where $\beta$ and $\eta$ are real constants, are proposed
phenomenologically to alleviate the coincidence problem and also
to prevent the universe to undergo the big rip \cite{pav}. To
study the evolution of the universe we are obliged to make use of
a specific interaction, which we choose a general form as
\cite{interaction1},
\begin{equation}\label{r4}
Q=3H(\tilde{\lambda}_m\rho_m+\lambda_d\rho_d)
\end{equation}
where $\tilde{\lambda}_m$ and $\lambda_d$ are real constants.
(\ref{r4}) reduces to (\ref{r2}) and (\ref{r3}) for
$\tilde{\lambda}_m=0$, $\lambda_d=\tilde{\lambda}_m$ and
$\lambda_d=0$ respectively. Note that (\ref{r4}) is the same as
the scalar
\begin{equation}\label{r5}
Q={1\over 3}V_{\mu}V_{\nu}V^{\gamma}
_{;\gamma}\left(\tilde{\lambda}_m T^{\mu \nu}_{(matter)}+\lambda_d
T^{\mu \nu}_{(dark)}\right),
\end{equation}
written in the comoving frame.  The equation of state (EoS)
parameter of dark energy, $w_d$, is defined by $P_d=w_d\rho_d$ and
the (EoS) parameter of matter, $0\leq w_m={P_m\over \rho_m}$, is
assumed to be a constant, e.g. for cold dark matter we have
$w_m=0$ and for radiation $w_m={1\over 3}$.

Different models have been proposed for dark energy, hereinafter we
adopt the (ECHDE) model \cite{wei} for which infrared cutoff is
taken as $L={1\over H}$. In this model, the dark energy density may
be expressed as
\begin{equation}\label{3}
\rho_d={3\over 8\pi}\left({c^2\over L^2}+{\alpha\over
L^4}\ln(L^2)+{\beta\over L^4}\right),
\end{equation}
$c$, $\alpha$ and $\beta$ are dimensionless real constants and
their values are still in debates in the literature \cite{wei}.

When ${\alpha\over L^4}\ln(L^2)+{\beta\over L^4}\ll {c^2\over
L^2}$, the model reduces to the ordinary holographic dark energy
model. The correction terms are relevant in the early universe,
and also in the late time provided that ${H^2\over {m_P^2}}$
becomes larger with respect to the present time. Note that in the
units used in this paper the reduced Planck mass is given by
$m_P=(8\pi)^{(-1/2)}$, so $c^2$ is the same as $n^2$ in
\cite{wei}. The ratio of dark energy density to critical density
is then
\begin{equation}\label{4}
\Omega_d=c^2-\alpha H^2\ln(H^2)+\beta H^2.
\end{equation}
By construction we must have
\begin{equation}\label{5}
0\leq c^2-\alpha H^2\ln(H^2)+\beta H^2\leq 1.
\end{equation}
Time derivative of $\Omega_d$ is obtained as
\begin{equation}\label{6}
\dot{\Omega}_d=-2\dot{H}H^{-1}(\alpha H^2-\Omega_d+c^2).
\end{equation}
In terms of the (EoS) of the universe,
\begin{equation}\label{7}
w={P_d+P_m\over {\rho_d+\rho_m}}=-1-{2\over 3}{\dot{H}\over H^2},
\end{equation}
(\ref{6}) can be written as
\begin{equation}\label{8}
\dot{\Omega}_d=3H(w+1)(\alpha H^2-\Omega_d+c^2).
\end{equation}
From (\ref{2}) one can derive
\begin{equation}\label{9}
\dot{r}=3Hr\left(w_d-w_m+(\lambda_d+r\tilde{\lambda}_m)\left({r+1\over
r}\right)\right),
\end{equation}
where $r={\rho_m\over \rho_d}$. Using $r={1-\Omega_d\over
\Omega_d}$, and
\begin{equation}\label{10}
w=w_d\Omega_d+w_m\Omega_m,
\end{equation}
we obtain
\begin{equation}\label{11}
w=-{1\over 3H}{\dot{\Omega_d}\over
1-\Omega_d}-{\lambda_d\Omega_d\over
(1-\Omega_d)}-\tilde{\lambda}_m+w_m.
\end{equation}

For $\alpha=\beta=0$, we have $\Omega_d=c^2$ and $w=w_m$. If
$w_m=0$ (e.g. when the matter is cold dark matter) and in the
absence of interaction we get $w=w_d=0$, implying that $\rho_d$ is
the same as dark matter. This was the motivation of taking another
infrared cutoff for the model in \cite{Li}. By taking the
interaction into account, the $\alpha=\beta=0$ model can describe
an accelerating universe with a non-dynamical $\Omega_d$
corresponding to a scaling solution.

From (\ref{8}) and (\ref{11})  we obtain
\begin{equation}\label{12}
w={(\tilde{\lambda}_m-\lambda_d-w_m+1)\Omega_d-\alpha
H^2-c^2-\tilde{\lambda}_m+w_m\over \alpha H^2-2\Omega_d+c^2+1},
\end{equation}
which results in
\begin{equation}\label{13}
\dot{H}={3H^2\over
2}{(\lambda_d-\lambda_m+1)\Omega_d+\lambda_m-1\over \alpha
H^2-2\Omega_d+c^2+1},
\end{equation}
where $\lambda_m=\tilde{\lambda}_m-w_m$.
 (\ref{13}) can be written as an autonomous first order
differential equation
\begin{equation}\label{14}
\dot{H}=G(H):={3\over 2}H^2{f(H)\over g(H)},
\end{equation}
where
\begin{eqnarray}\label{15}
f(H)&=&(\lambda_m-1)+(\lambda_d-\lambda_m+1)\left(c^2+\beta
H^2-\alpha
H^2ln(H^2)\right)\nonumber \\
g(H)&=& -c^2+1+(\alpha-2\beta)H^2+2\alpha H^2 ln(H^2).
\end{eqnarray}
Note that equation (\ref{14}) requires that $\dot{H}$ and all of
the higher order time derivatives of $H$ be zero at the time when
$\dot{H}=0$, i.e. $\dot{H}=0$ can occur only asymptotically.  As a
result $H$ can not cross $H=0$ (note that at $H=0$, $\dot{H}=0$).
So we may assume $H(t)>0$, $\forall t$ in the domain of validity
of our model. We also assume that $H$ is differentiable, therefore
$g(H)\neq 0$ and the sign of $g(H)$ does not change.

\section{Classification of the Hubble parameter behaviors}

Obtaining an analytical general solution to (\ref{14}) is not
possible. In this part, instead of solving (\ref{14}), by using
some mathematical methods based on the properties of Lambert
functions, we discuss and classify different possible behaviors of
the Hubble parameter dictated by this model, in terms of its
parameters $(\lambda_m, \lambda_d, c^2, \alpha, \beta)$. As we do
not fix the parameters, various behaviors for the model are
derived. Note that the results obtained in this section are
general and do not necessarily match with our present or early
universe. We will discuss this issue in the next section where as
we will see only some cases in the classification have necessary
(although not sufficient) conditions to describe the early and
late time acceleration of our universe.

For $\dot{H}+H^2>(<)0$ the universe is accelerating
(decelerating). For $\dot{H}>0$ the universe is in super-
acceleration (phantom) phase and for $\dot{H}<0$ the accelerated
universe is in the quintessence phase. Following the discussion in
the last part of the previous section, the sign of $\dot{H}$ is
unchangeable, hence the system is still in quintessence or phantom
phase and quintessence-phantom crossing does not occur.

By defining $u=H^2$, we have
\begin{equation}\label{122}
\dot{H}+H^2={F(u)\over g(u)}u,
\end{equation}
where $F(u):=A+Bu+Cu\ln(u)$, and
\begin{eqnarray}\label{123}
A&=&2(1-c^2)+3p\nonumber \\
B&=&2(\alpha-2\beta)+3q\nonumber\\
C&=&4\alpha+3s.
\end{eqnarray}
We have defined
\begin{eqnarray}\label{16}
p&=&\lambda_m-1+(\lambda_d-\lambda_m+1)c^2\nonumber\\
q&=&(\lambda_d-\lambda_m+1)\beta \nonumber\\
s&=&-(\lambda_d-\lambda_m+1)\alpha.
\end{eqnarray}
So $f$ in (\ref{15}) can be written as $f(u)=p+qu+su\ln(u)$.

To obtain the number of critical (fixed) points of the equation
(\ref{14}), and to get some insights about the behavior of the
system, we must find the number of zeroes and the behavior of
$f(H)$ in terms of the parameters of the model. In the same
manner, we must also study the behavior of $F(u)$. $f$ and $g$ in
(\ref{15}) and $F$ in (\ref{122}) have the functional form
$K(u):=a+bu+cu\ln(u)$. The general behavior of $K$ in terms of its
parameters is discussed in detail in the appendix.

$G(H)$ (or $\dot{H}$) in (\ref{14}) has a zero at $H=0$ and at
most two other zeroes at $H_1=\sqrt{u_1}$ and $H_2=\sqrt{u_2}$
determined in terms of Lambert W function: $W(x)$ \cite{Lambert},
as follows (the real branches of $W(x)$ are denoted by $W_0$ and
$W_{-1}$):
\begin{eqnarray}\label{17}
u_1&=&\exp\left(W_{-1}\left(-{p\over s}\exp{\left(q\over
s\right)}\right)-{q\over s}\right)\nonumber \\
u_2&=&\exp\left(W_{0}\left(-{p\over s}\exp{\left(q\over
s\right)}\right)-{q\over s}\right).
\end{eqnarray}
These are fixed points of the equation (\ref{14}), hence they can
not be crossed. We consider three domains: $D_1=(0,H_1)$,
$D_2=(H_1,H_2)$ and $D_3=(H_2,\infty)$. If $\exists t$, such that
$H(t)\in D_i$ then $H(t)$ will restricted to $D_i$. The sign of
$\dot{H}$ depends on the sign of $G$ in $D_i$, if $G(H)<(>)0$ for
$H\in D_i$, then $\dot{H}<(>)0$. The stability of the model at
critical points depends also on the sign of $g(H)$ in the region
where $H$ is restricted. We assume that $H$ is differentiable
(note that by construction $g(H)$ is well defined: $\forall H\in
\Re, g(H)\in \Re$). Hence $g(H)$ has no roots in the region where
$H$ is restricted, and therefore its sign does not change. For
simplicity we only study the cases with $g(H)>0$. The cases with
$g(H)<0$ can be treated in the same manner. Indeed, ${dG\over
dH}(H_i)<0$ implies that $\dot{H}>0$. This requires that ${df\over
dH}(H_i)$ and $g(H_i)$ have different signs ($H_i$ is the critical
point). For a $g(H_i)$ with opposite sign, $\dot{H}<0$. Note that
$g(H)>0$ is the only physical choice in regions including $H=0$:
$\lim_{H\to 0}g(H)=1-\Omega_d(H=0)>0$.

Using the results obtained in the appendix and the above
arguments, we can classify behaviors of the Hubble parameter as
follows:

For the very special case $s=0$, $\alpha\neq 0$ implies
$\lambda_d-\lambda_m+1=0$. In this situation the only fixed point
of the autonomous differential equation (\ref{14}) is $H=0$. If
$\lambda_m>1$, then $\dot{H}>0$ which gives rise to a super
accelerated expansion. For $\lambda_m<1$, $H=0$ becomes stable
fixed point and for $\lambda_m=1$ we obtain a de Sitter space time
$\dot{H}=0$.

According to the appendix we have the following possibilities (in
situations where $H_1=H_2$ we denote the root of $f$ by $H_1$):

 (I,1)$\left( s>0, p>0 , q> -s\left(\ln({p\over
s})+1\right)\right)$: We have $\dot{H}(t)>0$, so $H =0$ is an
unstable critical point and $\lim_{t\to \infty}H=\infty$.

(I,2) $\left( s>0, p>0 , q= -s\left(\ln({p\over
s})+1\right)\right)$: If $H(t)\in (0, H_1)$, then $\dot{H}(t)> 0$
and  $\lim_{t\to \infty}H(t)=H_1$. If $H(t)> H_1$, $\dot{H}(t)> 0$
and $\lim_{t\to \infty}H=\infty$.

So (I,1) and (I,2)  indicate that the expansion of the universe is
super-accelerated.

 (I,3)$\left( s>0, p>0, q<-s\left(\ln({p\over
s})+1\right)\right)$: If $H(t)\in(0,H_1)$, then $\dot{H}(t)>0$
corresponding the super-acceleration. For $H(t)\in (H_1,H_2)$, we
have $\dot{H}(t)<0$, which does not necessitate the acceleration
of the expansion. For $H(t)\in(H_2,\infty)$, $\dot{H}(t)>0$. Here
$H_1$ is a stable critical point.

 (I,4) $\left(s>0, p\leq 0\right)$: If $H(t)\in(0,H_1)$,
then $\dot{H}(t)<0$. For $H(t)\in(H_1,\infty)$ we have
$\dot{H}(t)>0$.

The same analysis can be done for (II) situations (see the
appendix):

(II,1)$\left(s<0, p<0, q< -s\left(\ln({p\over
s})+1\right)\right)$:  We have $\dot{H}(t)<0$ so $\lim_{t\to
\infty}H(t)=0$.

(II,2)$\left(s<0, p<0, q= -s\left(\ln({p\over
s})+1\right)\right)$: If $H(t)\in (0, H_1)$, then $\dot{H}(t)< 0$
and $\lim_{t\to \infty}H(t)= 0$. If $H(t)>H_1$, then
$\dot{H}(t)<0$ and $\lim_{t\to \infty}H(t)= H_1$.

(II,3) $\left( s<0, p<0, q>-s\left(\ln({p\over
s})+1\right)\right)$: If $H(t)\in(0,H_1)$, then $\dot{H}(t)<0$.
For $H(t)\in (H_1,H_2)$, we have $\dot{H}(t)>0$ and finally if
$H(t)\in(H_2,\infty)$, then $\dot{H}(t)<0$. In this situation
$H_2$ is a stable critical point.

(II,4) $\left(s<0, p\geq 0\right)$: If $H(t)\in(0,H_1)$ then
$\dot{H}(t)>0$. But if $H(t)\in(H_1,\infty)$ then $\dot{H}(t)<0$.
Here $H_1$ is a stable critical point.

In the above,  situations corresponding to $\dot{H}>0$ are related
to the phantom phase. We remind that in an open set $(0,H_1)$,
$(H_1, H_2)$ or $(H_2,\infty)$, the sign of $\dot{H}$ does not
change and, e.g. if the universe is in phantom phase $(\dot{H}>0)$
in some time of inflation, it will remain in this phase as long as
the model is valid. The same is true for $\dot{H}<0$. Note that
the cases corresponding to $\dot{H}<0$ do not necessitate an
accelerated expansion. In this case to see whether there is an
acceleration phase we must study $\dot{H}+H^2$.

So let us examine $\dot{H}+H^2$. This expression, besides at
$H=0$, has at most two positive zeroes at $H_3$ and $H_4$:

\begin{eqnarray}\label{1000}
H_3^2&=&\exp\left(W_{-1}\left(-{A\over C}\exp{\left(B\over
C\right)}\right)-{B\over C}\right)\nonumber \\
H_4^2&=&\exp\left(W_{0}\left(-{A\over C}\exp{\left(B\over
C\right)}\right)-{B\over C}\right).
\end{eqnarray}
As these are not fixed points, they can be crossed allowing
consecutive acceleration deceleration phases, but note that these
transitions may only occur when $\dot{H}<0$.

Again we can classify generally the model as follows (to avoid any
confusion, we use $III$ and $IV$ instead of $I$ and $II$ used
above, but note that $(III,i)$, and $(IV,i)$, again,  correspond
to $(I,i)$ and $(II,i)$ in the appendix respectively, and in
situations where $H_3=H_4$ we denote the root by $H_3$):

 (III,1) and (III,2)$\left(C>0, A>0 , B\geq -C\left(\ln({A\over
C})+1\right)\right)$: We have $\dot{H}+H^2(t)>0$.

(III,3)$\left(C>0, A>0, B<-C\left(\ln({A\over
C})+1\right)\right)$: If $H(t)\in(0,H_3)$, then
$\dot{H}+H^2(t)>0$. For $H(t)\in (H_3,H_4)$, we have
$\dot{H}+H^2(t)<0$. For $H(t)\in(H_4,\infty)$, $\dot{H}+H^2(t)>0$
holds.

(III,4)$\left( C>0, A\leq 0\right)$: If $H(t)\in(0,H_3)$, then
$\dot{H}+H^2<0$ and for $H(t)>H_3$: $\dot{H}+H^2>0$.

The same analysis can be done for (IV) situations (corresponding
to (II) in the appendix)

(IV,1) and (IV,2) $\left( C<0, A<0, B\leq -C\left(\ln({A\over
C})+1\right)\right)$: We have $\dot{H}+H^2<0$.

(IV,3)$\left(C<0, A<0, B>-C\left(\ln({A\over C})+1\right)\right)$:
If $H(t)\in(0,H_3)$, then $\dot{H}+H^2<0$. For $H(t)\in
(H_3,H_4)$, we have $\dot{H}+H^2>0$ and finally if
$H(t)\in(H_4,\infty)$, then $\dot{H}+H^2<0$.

(IV,4) $\left(C<0, A\geq 0\right)$: If $H(t)\in(0,H_3)$ then
$\dot{H}+H^2>0$. But if $H(t)\in(H_3,\infty)$ then
$\dot{H}+H^2<0$.

In summary by studying all various possible behaviors of the
Hubble parameter as the solution of autonomous differential
equation (\ref{14}), in (I-II, 1-4) the conditions leading to
$\dot{H}>0$ and $\dot{H}<0$ and all possible late time behaviors
of $H$ are specified. In this model the transition from phantom to
quintessence and vice versa are forbidden. This is due to the fact
that the points where $\dot{H}=0$, are critical points of the
theory. Although the conditions to have a phantom phase
$\dot{H}>0$ can be read from (I-II,1-4) but they don't elucidate
the conditions required to have the quintessence phase. So by
studying all possible behaviors of $\dot{H}+H^2$, conditions
needed for accelerated and decelerated expansion were specified.

\section{Physical Discussion and results}

In this part we try to study physical implications of our model in
the late time and separately in the inflation era of our universe.

\subsection{Late time acceleration}

From I and II,  we find out that all the possible late time
solutions in this model are de Sitter space-time and the
asymptotic values of the Hubble parameter may be the critical
points: $H=0$, $H=H_1$ or $H=H_2$ specified by (\ref{17}). We
remind that the system is restricted to the regions
$D_i=(H_i,H_j)$ bounded by critical points and the sign of
$\dot{H}$ is unchangeable in $D_i$. So if $\dot{H}<(>)0$ in $D_i$,
then $\lim_{t\to \infty}H=H_i(H_j)$. The model is stable at $H_i$
where $f(H_i)=0$ provided that ${1\over g(H_i)}{df\over
dH}(H_i)<0$, and a necessary condition for the system to tend to
$H=0$ is ${f(0)\over g(0)}<0$. For $\lim_{t\to \infty}H=0$, we
have ${dG\over dH}(0)=0$ and ${d^2G\over
dH^2}=3{{\lambda_m-1+(\lambda_d-\lambda_m+1)c^2}\over 1-c^2}<0$.
In the absence of interaction this inequality is satisfied but the
interaction may prevent the model to go asymptotically to $H=0$,
and instead, forces it to tend to $H_1$.

In the present era we have ${H^2\over {m_P^2}}\ll 1 $, therefore
${\alpha\over L^4}\ln(L^2)+{\beta\over L^4}\ll {c^2\over L^2}$ and
the correction terms have not important role. So at first sight,
it seems that the corrections are not relevant in the late time
acceleration, but this is not true. To emphasize this via a simple
example, let us take the ordinary holographic dark energy model.
In the absence (or smallness) of the correction terms, the only
critical point of (\ref{14}), is $H=0$. If we ignore the
interaction, we have $\dot{H}=-{3\over 2}H^2<0$ and $\lim_{t\to
\infty}H(t)=0$ and the universe tends to a static space time at
late time, besides, $\dot{H}+H^2<0$ and the acceleration does not
occur. During this evolution $\Omega_d$ has the constant value
$\Omega_d=c^2$. Here the correction terms will be also irrelevant
in the future evolution of the universe. Now let us take the
interaction into account and let $\left((\lambda_m-1)+{\lambda_d
c^2\over 1-c^2}\right)<0$ (this corresponds to $p\geq 0$ cases in
(I) and (II). Then $\dot{H}>0$ and the correction terms will
become relevant at late time.

As in our epoch ${H^2\over {m_P^2}}\ll 1 $, the present time
belongs to the region $(0,H_1)$ and if $\dot{H}<0$, then there is
no need to correction terms to study the future evolution of the
universe and we have $\lim_{t\to \infty}H(t)=0$, so the evolution
of the universe can be explained in the same manner as
\cite{pav1}. If, instead, $\dot{H}>0$ which is related to (I,1),
(I,2), (I,3) or (II,4) the correction terms become important and
the universe tends to a de Sitter space time characterized by :
$\lim_{t\to \infty}H(t)=H_1$. (I,1) implies $\lim_{t\to
\infty}H=\infty$ and is excluded by $\ref{5}$. So among various
cases in our classification in the previous section we are left
only with $s>0,\,\, p>0,\,\, q\leq -s\left(\ln({p\over
s}+1)\right)$ and the region $(0,H_1)$.

The value of $\Omega_d$ at late time is
\begin{equation}\label{2000}
\lim_{t\to \infty}\Omega_d={{1-\lambda_m}\over
{\lambda_d-\lambda_m+1}}.
\end{equation}
So the interaction, besides preventing the model to tend to $H=0$,
via the energy exchange, alleviates the coincidence problem. In
this case it is worth noting that if $\alpha=\beta=0$, then
$\lim_{t\to \infty}H(t)=\infty$. So the correction terms prevent
the Hubble parameter to become very large asymptotically.

\subsection{Inflation}
Among various possible acceleration phases, it seems that the
phantom phases, reported through the situations I and II, are not
consistent with a non eternal inflationary phase. Indeed as long
as the universe is dominated by (ECHDE) this inflationary phase
continues, i.e. the system is restricted to the domain specified
by critical points. So if the phantom phase occurs, can not be
ceased. Although crossing the critical points, $H_1$ and $H_2$, is
not possible but the system can cross $H_3$ and $H_4$ and
therefore transition from acceleration to deceleration and its
inverse are possible. To see this, as an example, let us take the
case (III,3) where for $H(t)\in (H_3,H_4)$, we have
$\dot{H}+H^2<0$ therefore $\dot{H}<0$ and $H$ is decreasing. But
$H_3$ is not a fixed point and the Hubble parameter crosses $H_3$
and enters in accelerating domain.

To describe the temporary inflationary phase, we must consider
$\dot{H}<0$  situations in I and II and then investigate the cases
reported in III and IV. If we expect that the inflation be ended,
we must select the situations allowing the transition from
acceleration to declaration phase. This is related to the presence
of the source term $Q$, allowing the energy exchange between dark
energy and matter. This is only allowed in (III,3) with
$H_{inf.}>H_4$, (III,4) with $H_{inf.}>H_3$, and (IV,3) with
$H_{inf.}\in (H_3,H_4)$, provided that the $H$ belongs to cases in
I and II where $\dot{H}<0$. $H_{inf}$ denotes the value of the
Hubble parameter during inflation. At the end of inflation, $H$ is
determined by $H_3$ (cases : (IV,3)and (III,4))or $H_4$ (case
(III,3)) determined by (\ref{1000}). The corresponding values of
$\Omega_d$ can then be read from (\ref{4}). In the above cases it
is straightforward to see that (IV) corresponds to an inflation
which is not past eternal.

\begin{figure}[h]
\centering
\includegraphics[angle=270,totalheight=2in]{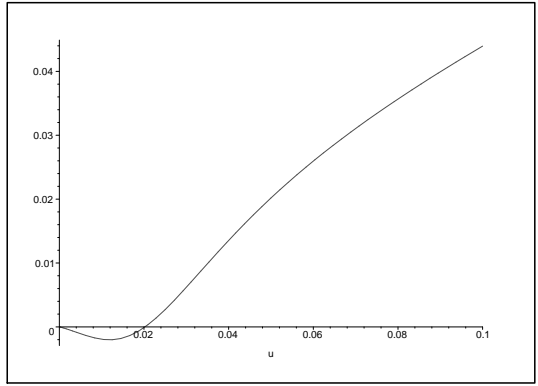}
\caption{$\dot{H}+H^2$ as a function of $u=H^2$, for
$\alpha=4.18,\,\, \beta=-10.71,\,\,\,\lambda_d=0.1, \,\,\,\,
\lambda_m =0, \,\,\,c^2=0.7$}\label{fig1}
\end{figure}

Until now, for the sake of generality we did not fix the values of
parameters $\{\alpha, \beta, c^2, \lambda_m, \lambda_d\}$. To be
more specific, and as an illustration of our results, let us take
the case (II,1) and (III,4) characterized by
\begin{eqnarray}\label{r8}
&&\lambda_d-\lambda_m+1)\left(\beta-\alpha\left(1+\ln({\lambda_m-1+(\lambda_d-\lambda_m+1)c^2\over
-\alpha(\lambda_m+\lambda_d+1)}\right)\right)<0\nonumber \\
&&(\lambda_m-\lambda_d-1)\alpha<0\nonumber\\
&&(\lambda_m-1)(1-c^2)+\lambda_dc^2<0
\end{eqnarray}
and
\begin{eqnarray}\label{r9}
&&(1-3\lambda_d+3\lambda_m)\alpha>0\nonumber \\
&&(3\lambda_d-3\lambda_m+1)c^2+3\lambda_m-1\leq 0,
\end{eqnarray}
respectively.  This situation describes a deceleration followed by
an acceleration phase as can be seen from the diagram of
$\dot{H}+H^2$ depicted in fig.(\ref{fig1}) for the optional choice
$\{\alpha=4.18, \beta=-10.71, c^2=0.7, \lambda_m=0, \lambda_d=0.1,
\}$ which satisfies(\ref{r8}) and (\ref{r9}) (note that the same
behavior is expected for all models whose parameters satisfy
(\ref{r8}) and (\ref{r9})). The inflation ends at $H^2=0.02
(\simeq 0.6 m_P^2)$ and (\ref{5}) restricts $H^2$ to $H^2<0.18
(0.95 m_P^2)$.

\begin{figure}[htbp]
\centering
\includegraphics[angle=270,totalheight=2in]{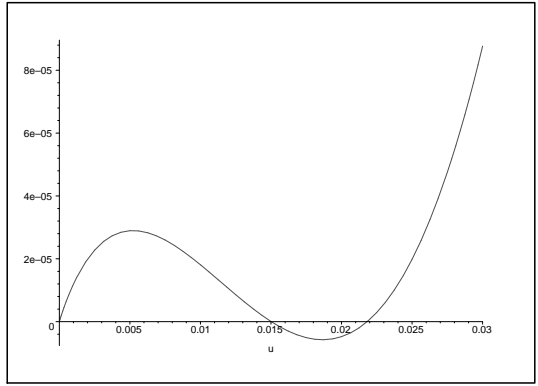}
\caption{$\dot{H}+H^2$ as a function of $u=H^2$, for
$\alpha=0.5,\,\, \beta=-0.5,\,\,\,\lambda_d=\lambda_m=0.103,
\,\,\,c^2=0.7$} \label{fig2}
\end{figure}

It is also interesting to see if the model permits to have
consecutive acceleration and deceleration phases. We remind that
 $\dot{H}+H^2$  has at most two zeroes besides $H=0$. So in
general,  besides acceleration-deceleration or
deceleration-acceleration transitions, it is also possible to have
successive deceleration-acceleration phases in this model.
Acceleration-deceleration-acceleration corresponds to (III,3) and
deceleration-acceleration-deceleration corresponds to (IV,3),
provided that the zeroes of $\dot{H}+H^2$ belongs to domain $D_i$
(specified by critical points) where the Hubble parameter is
restricted. Now to illustrate this result let us take the case
(II,1) and (III,3) characterized by (\ref{r8}) and :
\begin{eqnarray}\label{r10}
&&\alpha(3\lambda_m-3\lambda_d+1)>0\nonumber \\
&&3\lambda_m-1+\left(3(\lambda_d-\lambda_m)+1\right)c^2>0\nonumber
\\
&&3(\lambda_m-\lambda_d+1)(\alpha-\beta)+2\beta<\alpha(3\lambda_d-3\lambda_m-1)\times\nonumber
\\
&&\ln{(3\lambda_m-1)(1-c^2)+3\lambda_d c^2\over
\alpha(3\lambda_m-3\lambda_d+1)}.
\end{eqnarray}
$\dot{H}+H^2$ is depicted in terms of $u=H^2$, for the choice
$\{\alpha=0.5,\,\, \beta=-0.5,\,\,\,\lambda_d=\lambda_m=0.103,
\,\,\,c^2=0.7\}$, which satisfy (\ref{r8}) and (\ref{r10}), in
fig.(\ref{fig2})(again note that the same qualitative behavior is
expected for all models whose parameters satisfy (\ref{r8}) and
(\ref{r10})).  The model has an acceleration phase for
$H^2>0.0218$ (or in $\hbar=c=1$ units $H^2>0.548 m_P^2$) a
deceleration phase for $0.0150(=0.377 m_P^2)<H^2<0.0218(=0.548
m_P^2)$ and again an acceleration phase for $0<H^2<0.0150(=0.377
m_P^2)$. Note that (\ref{5}) restricts the model to
$H^2<1.19=(29.9 m_P^2)$. Models with double inflation (i.e. two
stages of inflation) were studied and reported before in the
literature \cite{turn}.

The scope of our discussion was only restricted to the study of
possible accelerations of the universe in early and late times.
However a realistic general model of inflation must also be
capable to describe physical problems such as the reheating and
the cosmological perturbations and so on.

In the inflationary era, the universe is dominated by $\rho_d$ for
which the pressure is negative and by giving rise to $\dot{H}+H^2>0$
drives the inflation. In different models, there may be different
scenarios for the reheating. As an example, in the scalar field
model and in the slow roll approximation, after the inflation the
scalar field decays to other particles during a rapid coherent
oscillation in the bottom of the potential slope \cite{Kolb}. In our
model and in the absence of scalar fields, there can be another
possibilities for reheating:

By taking $w_m=1/3$ and $\tilde{\lambda}_m=0$, the model acts
similar to warm inflation models where the inflaton decays to
radiation (ultra-relativistic particles) during inflation
\cite{warm}. This is due to the source term, $3H\lambda_d \rho_d$,
which allows energy exchange between these components. In the
inflationary era, the universe is dominated by $\rho_d$, and this
source term is significant in (\ref{2}). In contrast, after the
inflation, the contribution of $\rho_m$ becomes more significant
and the universe becomes radiation dominated for
$\dot{H}+2H^2\approx 0$ which occurs in deceleration era. In this
era the radiation creation from $\rho_d$ is not significant. Note
that the radiation dominated era occurs for $\dot{H}+2H^2\approx
0$. The corresponding value of the Hubble parameter can be
obtained by solving the equation $A_r+B_ru+C_ru\ln(u)=0$, where
$A_r=4(1-c^2)+3p$; $B_r=4(\alpha-2\beta)+3q$ and $C_r=8\alpha+3s$.

One can also assume another possibility for reheating by taking
$w_m\neq {1\over 3}$. This assumption may be valid provided that
$\rho_m$ decays to ultra-relativistic particles at the end of
inflation, giving rise to the preheating or the reheating. However
the study of this era requires that we consider the contribution
of baryonic matter in (\ref{2}) and add the corresponding
interaction terms to our equations.

At the end it is worth to note that the above conditions posed on
the parameters of the model, $(\alpha, \beta, \lambda_d,
\lambda_m)$, in (III,3),(III,4), and (IV,3) although are necessary
conditions for transient inflation but are not sufficient. For
example for a given model with specified parameters the e-folds
number $N:=\ln\left({a(t_{end})\over a(t_i)}\right)$ must be
calculated. It is given by
\begin{equation}\label{int}
N={1\over 2}\int_{u_{end}}^{u_i} {du\over G(u)},
\end{equation}
where as before $u:=H^2$, $i$ and $end$ denote the beginning and
the end of the inflation and $G(u)$ is given by (\ref{14}). In the
case (III,4); (III,3); and (IV,3) we have $u_{end}=H_3^2$;
$u_{end}=H_4^2$; and $u_{end}=H_3^2$ respectively. In the case
(IV,3), $H_3^2<u_{i}\leq H_4^2$. Note that in other two cases
$u_i\in (u_{end},V)$ where $V$ is the maximum value of $u$
satisfying $(\ref{5})$. For a viable model we must have $60\leq
N$. This put more constraints on the parameter of the model.
However for a general model, analytically solving (\ref{int}) or
obtaining a lower bound for it, are very complicated tasks.

\section{Conclusion}
In this paper we considered  a spatially flat FRW universe
dominated by (ECHDE) and a barotopic matter interacting via a more
general source term with respect to other papers in this subject.
We took the apparent horizon as the infrared cutoff (this is the
more natural choice adopted in the literature). Considering the
effects of thermal fluctuations around equilibrium, quantum
fluctuations, or charge and mass fluctuations, modifies the
entropy attributed to the horizon. We applied these corrections to
the apparent horizon entropy and achieved to obtain an autonomous
differential equation for the Hubble parameter. Then we obtained
the critical points of the model and classified the behavior of
the system in terms of the parameters of the interaction and
(ECDHE). For this purpose we used algebraic features of the
autonomous differential equation and $LambertW$ functions.

Although in the present time the corrections are marginal but they
may play an important role in the early and late times. We deduced
that the correction terms may force the universe to tend to a de
Sitter space-time at late time. We obtained the possible ultimate
value of the Hubble parameter and also derived the corresponding
dark energy density. We showed that the coincidence problem is
alleviated in this model.

In addition we studied some necessary (although not sufficient)
conditions for the model to describe the acceleration phase in
inflation era. The inflation was assumed to be transient and the
possible values of the Hubble parameter at the end of inflation
were derived. \vspace{1cm}

{\bf{Appendix}}\vspace{0.5cm}

In this part we study the behavior and the properties of the roots
of $K(u):=a+bu+cu\ln(u)$.

${dK(u)\over du}$ has and only has a root at
$\tilde{u}=\exp\left(-{b+c\over c}\right)$, so following the
Rolle's theorem $K(u)$ has at most two positive roots which we
denote $u_1$ and $u_2>u_1$.

We have also ${d^2K\over du^2}(\tilde{u})={c\over \tilde{u}}$.
Generally our model can be classified into three classes: $c=0$,
I: $c>0$ and II: $c<0$.

In the case I:

1) for $K(0)>0$ and $K(\tilde{u})>0$, $K$ has no roots and $K(u)$
is always positive.

2) For $K(0)>0$ and $K(\tilde{u})=0$, $K$ has only one root and
$K(u)$ is positive.

3) For $K(0)> 0$, and $K(\tilde{u})<0$, $K$ has two roots: $0<
u_1<u_2$, and $K(u>u_2)>0$, $K(u_1<u<u_2)<0$ and $K(0<u<u_1)>0$.
In this case ${dK\over du}(u_1)<0$ and ${dK\over du}(u_2)>0$.

4) For $K(0)\leq 0$ and $K(\tilde{u})<0$, $K$ has only one non
zero root $u_1$. We have also $K(u>u_1)>0$, $K(0<u<u_1)<0$ and
${dK\over du}(u_1)>0$.

The cases (I,1), (I,2), (I,3), and (I,4) correspond to
\begin{eqnarray}
&&c>0, a>0, b>-c\left(\ln{a\over c}+1\right);\nonumber \\
&&c>0, a>0, b=-c\left(\ln{a\over c}+1\right);\nonumber\\
&&c>0, a>0, b<-c\left(\ln{a\over c}+1\right), and;\nonumber\\
&&c>0, a\leq 0,
\end{eqnarray}
respectively.

In the case II:

1) for $K(0)<0$ and $K(\tilde{u})<0$, $K$ has no roots and $K(u)$
is always negative.

2) For $K(0)<0$ and $K(\tilde{u})=0$, $K$ has only one root and
$K(u)$ is always negative.

3) For $K(0)< 0$, and $K(\tilde{u})>0$, $K$ has two roots:
$0<u_1<u_2$, and $K(u>u_2)<0$, $K(u_1<u<u_2)>0$ and
$K(0<u<u_1)<0$. In this case ${dK\over du}(u_1)>0$ and ${dK\over
du}(u_2)<0$.

4) For $K(0)\geq 0$ and $K(\tilde{u})>0$, $K$ may have only one
nonzero root $u_1$, and $K(u>u_1)<0$, $K(0<u<u_1)>0$ and ${dK\over
du}(u_1)<0$.

The cases (II,1), (II,2), (II,3), and (II,4) correspond to
\begin{eqnarray}
&&c<0, a<0, b<-c\left(\ln{a\over c}+1\right);\nonumber \\
&&c<0, a<0, b=-c\left(\ln{a\over c}+1\right); \nonumber \\
&&c<0, a<0, b>-c\left(\ln{a\over c}+1\right), and; \nonumber \\
&&c<0, a\geq 0
\end{eqnarray}
respectively.

The numbers of the roots of $K$ can also be explained in terms of
Lambert W function: $W(x)$ \cite{Lambert}, in a more
straightforward way. The real branches of $W(x)$ are denoted by
$W_0$ and $W_{-1}$. For real $x$, if $-{1\over e}<x<0$ there are
two possible real values for $W(x)$: $-1< W_0(x)$, and
$W_{-1}(x)<-1$. We have also $W_0(-{1\over e})=W_{-1}(-{1\over
e})=-1$.

For ${a\over c}e^{b\over c}\leq 0$ (the cases 4 in I and II), the
solution of the equation $K=0$ in terms of Lambert W function is
\begin{equation}
u=\exp\left(W_0\left(-{a\over c}\exp{\left(b\over
c\right)}\right)-{b\over c}\right)
\end{equation}
For ${1\over e}<{a\over c}e^{b\over c}$, $K$ has no real roots
(the cases 1 in I and II), and for $0<{a\over c}e^{b\over
c}<{1\over e}$, $K$ has two roots (the cases 3 in I and II):
\begin{eqnarray}
u_1&=&\exp\left(W_{-1}\left(-{a\over c}\exp{\left(b\over
c\right)}\right)-{b\over c}\right)\nonumber \\
u_2&=&\exp\left(W_{0}\left(-{a\over c}\exp{\left(b\over
c\right)}\right)-{b\over c}\right).
\end{eqnarray}
For ${1\over e}={a\over c}e^{b\over c}$, $u_1=u_2$ corresponding
to the cases 2 in I and II.

\vspace{2cm}

{\bf Acknowledgments}

H. Mohseni Sadjadi would like to thank the Center of Excellence on
the Structure of Matter of the University of Tehran for its
supports. We would deeply thank the referee for his critical
comments on this paper.

\end{document}